\documentclass[doublecol]{epl2}
\usepackage{graphicx}
\usepackage{color}
\usepackage{amsmath, amsthm, amssymb}
\newcommand{\be}{\begin{equation}}
\newcommand{\ee}{\end{equation}}
\newcommand{\bea}{\begin{eqnarray}}
\newcommand{\eea}{\end{eqnarray}}

\begin{document}

\title{Phase transitions in systems of hard rectangles with non-integer aspect ratio}
\shorttitle{Hard rectangle gas}
\author{Joyjit Kundu\and R. Rajesh}
\author{Joyjit Kundu\footnote{joyjit@imsc.res.in}
and R. Rajesh\footnote{rrajesh.imsc.res.in} }
%\email{joyjit@imsc.res.in}
%\institute{The Institute of Mathematical Sciences, C.I.T. Campus,
%Taramani, Chennai 600113, India} 
%\author{R. Rajesh}
\shortauthor{Joyjit Kundu \etal}
%\email{rrajesh@imsc.res.in}
\institute{The Institute of Mathematical Sciences, C.I.T. Campus,
Taramani, Chennai 600113, India}

\date{\today}

\abstract 
{We investigate, using Monte Carlo simulations, the phase diagram of a 
system of hard rectangles of size $m\times mk$ on a square lattice 
when the aspect ratio $k$ is a non-integer. The existence 
of a disordered isotropic phase, a nematic with only orientational order, a columnar phase with 
orientational and partial translational order, and a high density phase 
with no orientational order is shown. The high density
phase is a solid-like sublattice phase only if the length
and width of the rectangles are not mutually prime, else, it is 
an isotropic phase. The minimum value
of $k$ beyond which the nematic and columnar phases exist are
determined for $m=2$ and $3$.  The nature of the
transitions between different phases is determined, and the critical
exponents are numerically obtained for the continuous transitions.}

\pacs{64.60.De}{{\bf Statistical mechanics of model systems (Ising 
model, Potts model, field-theory models, Monte Carlo techniques, etc.)}} 
\pacs{05.50.+q}{{\bf Lattice theory and statistics (Ising, Potts, 
etc.)}} 
\pacs{64.70.mf}{{\bf Theory and modeling of specific liquid 
crystal transitions, including computer simulation}}

\maketitle

Understanding the nature of the 
different phases and the transitions between them in a system of hard 
rods has significance for more complex physical systems such as liquid 
crystals~\cite{degennesBook}, tobacco mosaic virus~\cite{wen1989}, $fd$ 
virus~\cite{eric2008,fraden1997,fraden2000}, silica 
colloids~\cite{jacs2011,kuijk2012}, boehmite 
particles~\cite{buining1993,kooij1996}, DNA origami 
nanoneedles~\cite{nano2014}, and adsorption of gas particles on metal 
surfaces~\cite{taylor1985,bak1985,rikvold1991,binder2000, evans2000}. In 
three dimensional continuum, the system of hard rods undergoes an 
entropy driven phase transition from a low density isotropic phase to a 
high density nematic phase that has orientational 
order~\cite{onsager1949,flory1956b,zwanzig1963}. Further increase in 
density may result in a smectic phase with orientational and partial 
translational order and a solid phase~\cite{frenkel1997}. In two 
dimensions, the system undergoes a Kosterlitz Thouless type phase 
transition from an isotropic phase to a power law correlated 
phase~\cite{thouless1973}. Hard cuboids on a cubic lattice and hard 
rectangles on a square lattice are the corresponding lattice analogues. 
While the complete phase diagram for cuboids is not known, the system of 
hard rectangles has a rich phase diagram~\cite{joyjit_rectangle}.

Consider a system of monodispersed hard rectangles of size $m \times m 
k$ ($k>1$) on a square lattice, interacting only through excluded volume 
interaction. When $m=1$ (hard rods), the system undergoes two 
transitions with increasing density for aspect ratio $k \geq 
7$: first from a low density disordered phase to an 
intermediate density nematic 
phase~\cite{ghosh2007,giuliani2013,fernandez2008a,fernandez2008b,fischer2009}, 
and second, from the nematic phase to a high density disordered 
phase~\cite{joyjit_dae,joyjit2013}. The high density disordered phase 
has been argued to be a reentrant low density disordered 
phase~\cite{joyjit_rltl2013}.  
When $k=2$ (dimers), it may be rigorously 
shown that there are no transitions at non-zero densities of 
vacancies~\cite{lieb1972,Gruber1971,Kunz1970,Heilmann1970}, though at 
full packing, the correlations decay algebraically on bipartite 
lattices~\cite{Stephenson1963}, and exponentially on non-bipartite 
lattices~\cite{sondhi2002}. For $k \gg1$, the existence of the nematic 
phase at intermediate densities may be proved 
rigorously~\cite{giuliani2013}. The only exact solutions that exist are 
for the model of hard rods on a Bethe-like 
lattice~\cite{dhar2011,joyjit_rltl2013}.

For $m \geq 2$ and integer $k$, four different phases have been 
observed: isotropic, nematic, columnar, and solid-like sublattice 
phases~\cite{joyjit_rectangle,joyjit_rec_asym,nkr14}. 
The nematic phase exists only for $k\geq 7$ for $m=2,3$. The columnar phase
exists only when $k \geq 4$ for $m=2$ and for $k\geq 2$ for $m=3$.
For large enough $k$, with increasing
density, the system transits successively from isotropic to nematic to 
columnar to sublattice phase. The nature
of the phase transitions has also been studied. The isotropic-nematic 
transition belongs to the Ising universality class for all $m$. When $m 
\geq 3$, all other transitions are first order in nature. When $m=2$, 
all transitions except the isotropic-columnar transition are continuous 
and belong to the Ising or Ashkin-Teller universality class. The 
isotropic-columnar transition is continuous for $k=5$ and first order 
for $k=6$, implying the existence of a tricritical point at an intermediate value of 
$k$~\cite{joyjit_rectangle}. 

All the above results are for integer values of the aspect ratio $k$. 
Within viral expansion and Bethe approximation, the results do not 
depend on whether $k$ is an integer or not~\cite{joyjit_rec_asym}. 
However, these results are presumably valid only for large $k$.
For smaller $k$, when the applicability of the model
for adsorption of gas particles on metal surfaces is more relevant,
it is not clear what the effect of $k$ being an non-integer is. 
What are the different phases and the phase diagram when $k$ is rational but not 
an integer? 
What are the minimum values of $k$ beyond which the
nematic and columnar phases exist? 
Answering these questions will allow us to obtain the complete phase diagram for 
the system of hard rectangles. In this paper, we obtain the phase 
diagram for $m=2$, when $k$ is a half-integer, using large scale Monte 
Carlo simulations. The isotropic-columnar
transition is shown to be discontinuous while the isotropic-nematic and 
nematic-columnar transitions are shown to be
continuous. The critical exponents are numerically 
estimated for the continuous transitions. We find that the columnar phase
exists only when $k \geq 11/2$ for $m=2$ and when $k \geq  13/3$ for $m=3$.
The nematic phase exists only when $k\geq 15/2$ for $m=2$ and when $k\geq 22/3$
for $m=3$.

\section{\label{sec:model_ch05}The Model and Monte Carlo algorithm}

Consider monodispersed hard rectangles of size $m\times mk$ on a square 
lattice of size $L\times L$ with periodic boundary conditions.  Each 
rectangle is oriented either horizontally or vertically. A horizontal 
(vertical) rectangle occupies $mk$ lattice sites along x (y)-direction 
and $m$ lattice site along y (x)-direction. Each site may have at most 
one rectangle passing through it. We associate an activity $z=e^{\mu}$ 
to each rectangle, where $\mu$ is the chemical potential. In this paper, 
we restrict the aspect ratio $k$ to non-integers.

We simulate the system in the constant $\mu$ grand canonical ensemble 
using an efficient algorithm involving cluster moves that has been shown 
to be very useful in equilibrating hard core systems of extended 
particles at high 
densities~\cite{joyjit_dae,joyjit2013,joyjit_rectangle}. Here we briefly 
review the algorithm. Starting from a valid configuration, a row or a 
column (say a row) is chosen at random. All horizontal rectangles whose 
bottom-left corners (heads) are on that row are evaporated, keeping the rest of 
the configuration unchanged. The row now consists of intervals of empty 
sites, separated by sites that are either occupied by rectangles or can 
not be occupied due to the hard core constraint. The empty intervals of 
the row are re-occupied by a new configuration of horizontal rectangles 
with the correct equilibrium grand canonical probabilities. The 
calculation of these probabilities reduces to a solvable one-dimensional 
problem. If a column is chosen, similar evaporation-deposition moves are 
performed for vertical rectangles. Equilibration is faster on including 
a flip move in which a square plaquette consisting of $\ell$ aligned 
horizontal or vertical rectangles is rotated by $\pi/2$, where $\ell$
is the ratio of the least common multiple of $m$ and $m k$ to $m$. A detailed 
description of the implementation of the algorithm for the system of 
hard rectangles is described in Ref.~\cite{joyjit_rectangle}. Unlike
for integer $k$, the flip move is less effective for non-integer $k$ because the 
rotatable plaquettes are larger in size and hence, have lower probability to occur
during the simulations. This makes it difficult to equilibrate the systems at high 
densities. Other 
implementations of the algorithm include lattice models of hard 
rods~\cite{joyjit_dae,joyjit2013}, hard discs~\cite{trisha_knn}, and 
mixtures of dimers and hard squares~\cite{kabir2014}.

\section{\label{sec:def_phases_ch05}Different Phases}

As for integer $k$, we observe four different phases in the simulations: 
an isotropic (I) phase, a nematic (N) phase, a columnar (C) phase and a 
high density (HD) phase. The I phase is disordered. In the N phase, rectangles orient preferably along the horizontal or 
vertical direction, but they do not have any positional order. Each row 
or column on an average contains equal number of heads of rectangles. 
The columnar (C) phase has orientational order and translational order 
only in the direction perpendicular to the nematic orientation. When 
$m=2$, in the columnar phase, if the majority of the rectangles are 
horizontal (vertical), their heads lie mostly on even rows (columns) or 
odd rows (columns). Hence, there are $4$ (in general $2 m$) symmetric C 
phases. The I, N and C phases are observed when $m\geq 2$, for both 
integer and non-integer $k$.

The HD phase has no orientational order. But it may or may not possess 
translational order depending on the length and width of the rectangles. 
Let, the greatest common divisor of the length and width be denoted by 
$p$. We divide the square lattice into $p^2$ sublattices by assigning to 
a site $(i,j)$ a label $(i \mod p) +p \times (j \mod p)$. In the fully 
packed limit, it is straightforward to verify that the heads of the 
rectangles occupy one of the $p^2$ sublattices. We expect this phase to 
be stable to introduction of vacancies at densities close to the full 
packing. If $p>1$, the HD phase is a sublattice phase with complete 
translational order but no orientational order~\cite{joyjit_rectangle}. 
On the other hand, when $p=1$ (length and 
width are mutually prime), the HD phase is disordered with 
no orientational or translational order. Since existing evidence for 
$m=1$ suggests that the high density disordered phase is qualitatively 
similar to the low density I phase~\cite{joyjit2013,joyjit_rltl2013}, 
we expect the same to hold for 
$m\geq 2$ whenever the HD phase is disordered.

\begin{figure}
\centering 
\includegraphics[width=\columnwidth]{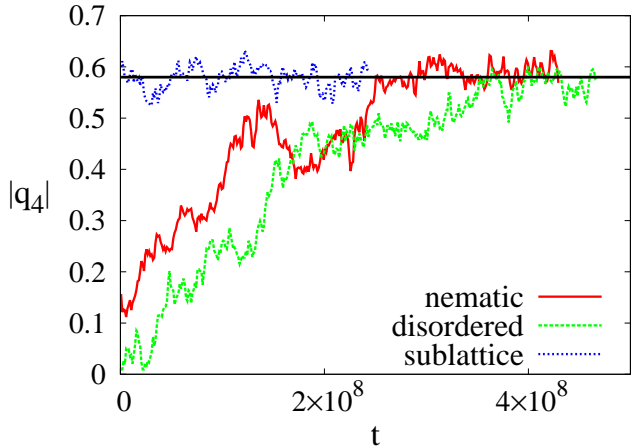}
\caption{Time evolution of the sublattice order parameter $q_1$, starting from 
three different types of initial configurations: nematic, 
disordered, and sublattice phases. The straight line is $|q_1|=0.580$.
The data are $\mu=10.8$ and for L=960. The equilibrium density is $ 
\approx 0.962$.}
\label{fig:w4k6} 
\end{figure}

When $m\geq 2$ and integer $k$, $p=m>1$ and the HD is 
known to be a sublattice phase~\cite{joyjit_rectangle}, consistent with 
the above argument. To further confirm that the HD phase is a sublattice 
phase when $p >1$, but $k$ is a non-integer, we simulate the system of 
rectangles of size $4\times 6$, for which $p=2$. We 
divide the lattice into $p^2=4$ sublattices. The 
sublattice order parameter is defined as $q_1=n_0-n_1-n_2 + n_3$, where 
$n_i$ is the fraction of sites occupied by rectangles whose heads
are on the $i$th sublattice. It is straightforward to check that 
$\langle |q_1| \rangle \neq 0$ only for the sublattice phase. To show the existence 
of sublattice phase at high density, a large value of $\mu$ 
is chosen ($\mu=10.8$), and the temporal evolution of $|q_1| $ is tracked, 
starting from three different initial configurations: 
nematic, disordered and sublattice phases (see Fig~\ref{fig:w4k6}). At 
large times, the system reaches a stationary state that is 
independent of the initial configuration, ensuring equilibrium. 
For this choice of $\mu$, the fraction of occupied sites $\rho$ fluctuates
around $0.962$. In 
equilibrium, $\langle | q_1| \rangle \approx 0.580$, clearly showing the existence of 
a sublattice phase. 

\section{\label{sec:phase_diag_ch05}Phase Diagram for $m=2$}

\begin{figure}
\centering 
\includegraphics[width=\columnwidth]{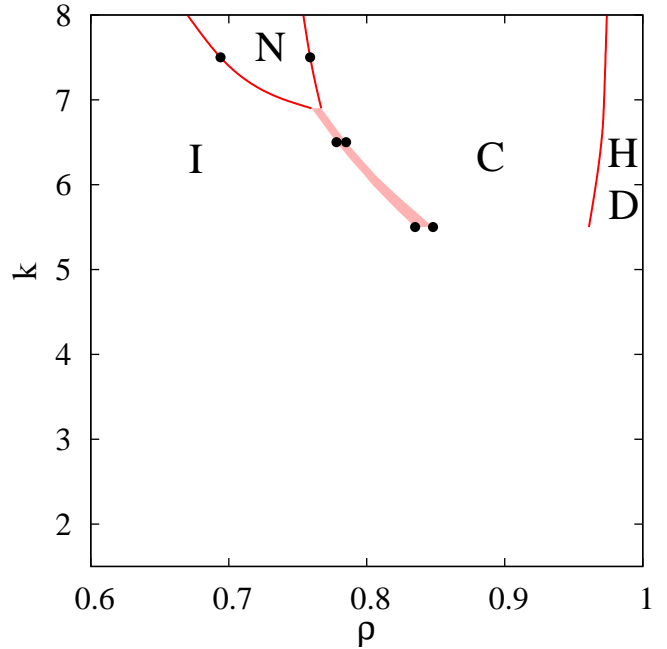}
\caption{Phase diagram for rectangles of size $2 \times 2 k$, where $k$ 
is restricted to non-integer values. I, N, C and HD denote isotropic, 
nematic, columnar and high density phases respectively. The data points 
are from simulation, while the continuous lines and shaded portions are 
guides to the eye. The shaded portion denotes regions of phase 
coexistence.}
\label{fig:phase_diag_m2_ch05} 
\end{figure}

The phase diagram for $m=2$ and non-integer $k$ 
is shown in Fig.~\ref{fig:phase_diag_m2_ch05}, where the data points are
obtained from Monte Carlo simulations and the lines, based on 
analysis of the phase diagram for large $k$~\cite{joyjit_rec_asym} 
are guides to the eye. The low density phase is an I phase for all $k$. Since the 
length and width of the
rectangles are mutually prime, 
$p=1$, and the HD phase is a reentrant I phase. No phase transitions are 
observed when $k \leq 9/2$. The C phase exists only for
$k \geq 11/2$, while the N phase exists only for $k \geq 15/2$. 

We could not numerically obtain any data point on the C-HD phase
boundary as it is not possible to equilibrate the systems within
available computer time at high
densities for $k \geq 11/2$. However, the critical density for
C-HD phase transition was argued to behave asymptotically as $1-a/(m k^2)$, for $k \gg
1$, where $a$ is a constant.
Likewise, for large $k$,
the critical density for the I-N phase transition scales as $A k^{-1}$,
where $A$ is a constant, independent of $m$, and 
that for the N-C phase transition tends to a
non-zero constant~\cite{joyjit_rectangle,joyjit_rec_asym}. The solid lines in
Fig.~\ref{fig:phase_diag_m2_ch05} follow these asymptotic behavior for large $k$.

The I-C transition is found to be first order for both 
$k=11/2$ and $13/2$. The shaded region in Fig.~\ref{fig:phase_diag_m2_ch05} denotes the region of phase 
coexistence at a first order phase transition. We find that the I-N
and N-C transitions are both continuous. These transitions are
analyzed in detail below.

\section{\label{sec:phase_m2_ch05}Critical behavior for $m = 2$}

We now study the nature of the different phase transitions for the system of $2 \times 2 k$ 
rectangles, where $k$ is half-integer. To study the I-N transition, we define the order parameter
\be
q_2 = n_h-n_v, \label{eq:m1_ch05}
\ee
where $n_h$ and $n_v$ are the fraction of sites occupied by the 
horizontal and vertical rectangles respectively. In the I phase $\langle 
|q_2| \rangle =0$, while in the N phase, $\langle |q_2| \rangle \neq 
0$.

The I-C and N-C phase transitions are best studied with the order parameter
\be
q_3= |n_{re}-n_{ro}| -|n_{ce}-n_{co}|, \label{eq:m2_ch05}
\ee
where $n_{re}$ ($n_{ro}$) is the fraction of sites occupied by 
rectangles whose heads are in the even (odd) rows, and $n_{ce}$ 
($n_{co}$) is the fraction of sites occupied by rectangles whose heads 
are in the even (odd) columns. In the I and N phases, $n_{re} \approx 
n_{ro}$, and $n_{ce} \approx n_{co}$, implying that $\langle |q_3| 
\rangle =0$.  In the C phase, either $n_{re} \neq n_{ro}$ and $n_{ce} 
\approx n_{co}$, or $n_{ce} \neq n_{co}$  and $n_{re} \approx n_{ro}$ 
implying that $\langle |q_3| \rangle \neq 0$. 

The other relevant thermodynamic quantities are the 
second moment $\chi_i$ and the Binder cumulant $U_i$, defined as
\begin{subequations}
\label{eq:thermo-definition_ch05}
\bea
U_i&=&1- \frac{\langle q_i^4 \rangle} {3 \langle q_i^2 \rangle ^2}.\\
\chi_i&=&\langle q_i^2 \rangle L^2, \label{eq:chi}
\label{eq:U}
\eea
\end{subequations}
where $i=2,3$.
Near the critical point, the singular behavior is captured by  finite-size scaling:
\begin{subequations}
\label{eq:scaling}
\bea
\langle |q_i |\rangle &\simeq& L^{-\beta/\nu} f_q(\epsilon L^{1/\nu}), 
\label{eq:Qscaling}\\ 
U_i &\simeq& f_u(\epsilon L^{1/\nu}), \label{eq:Uscaling}\\
\chi_i & \simeq& L^{\gamma/\nu} f_{\chi}(\epsilon L^{1/\nu}),
\label{eq:chiscaling}
\eea
\end{subequations}
where $\epsilon=(\mu-\mu_c)/\mu_c$, where $\mu_c$ is the critical 
chemical, $\beta$, $\gamma$, $\nu$ are the critical exponents, and 
$f_q$, $f_u$, and $f_{\chi}$ are scaling functions.

\subsection{\label{sec:IN_ch05}Isotropic--Nematic (I-N) transition}

\begin{figure}
\includegraphics[width=\columnwidth]{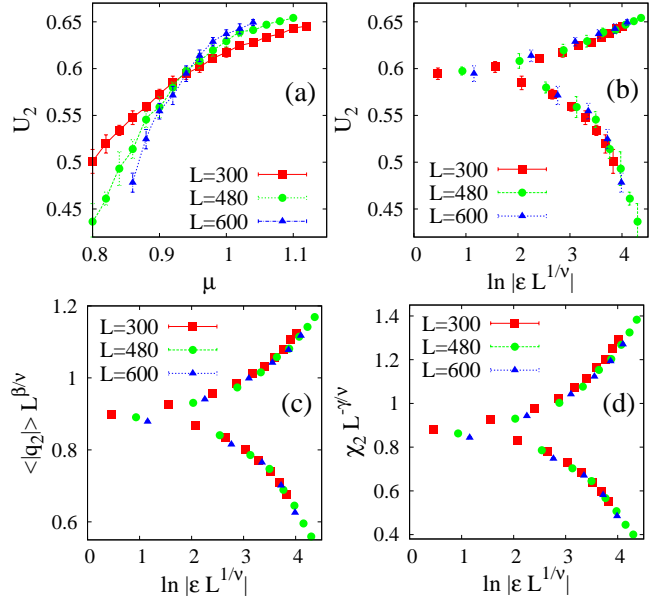}
\caption{The critical behavior near the I-N transition for rectangles of 
size $2 \times 15$ ($k=15/2$). (a) The data for Binder cumulant for 
different system sizes intersect at $\mu_{I-N}^c \approx 0.945$ 
($\rho^c_{I-N}$ $\approx 0.694$). The data for different $L$ near the 
I-N transition for (b) Binder cumulant, (c) order parameter, and (d) 
second moment of the order parameter collapse onto a single curve when 
scaled as in Eq.~(\ref{eq:scaling}) with the Ising exponents 
$\beta/\nu=1/8$, $\gamma/\nu=7/4$, and $\nu = 1 $.}
\label{fig:scaling_IN_ch05}
\end{figure}

We study the I-N transition for $2\times 15$ ($k=15/2$) 
rectangles using the order parameter $q_2$. Since the 
N phase may have orientational order only in the  horizontal 
or vertical direction, we expect the I-N transition to be in the two-dimensional Ising 
universality class, as has been confirmed for integer $k$, 
when $m=1$~\cite{fernandez2008a} and $m=2,3$~\cite{joyjit_rectangle}, and for systems of polydispersed rods~\cite{velenik2006,rs14}.
The data for $U_2$ for different system 
sizes intersect at $\mu=\mu_{I-N}^c\approx 0.945$ [see 
Fig.~\ref{fig:scaling_IN_ch05}(a)]. The corresponding critical density is $\rho_{I-N}^c 
\approx 0.694$, which is less than $\rho_{I-N}^c \approx 0.745$ for 
$k=7$~\cite{joyjit2013}, consistent with $\rho_{I-N}^c \approx 
A k^{-1}$~\cite{joyjit_rec_asym}. The data for $U_2$ [see 
Fig.~\ref{fig:scaling_IN_ch05}(b)], $\langle |q_2| \rangle$ [see 
Fig.~\ref{fig:scaling_IN_ch05}(c)], and $\chi_2$ [see 
Fig.~\ref{fig:scaling_IN_ch05}(d)] for different system sizes collapse 
onto a single curve when scaled as in 
Eq.~(\ref{eq:scaling}) with Ising exponents 
$\beta/\nu=1/8$, $\gamma/\nu=7/4$, and $\nu = 1 $. For larger values of 
$k$, integer or otherwise, we expect the I-N transition to be in the 
Ising universality class.

\subsection{\label{sec:NC_ch05}Nematic--Columnar (N-C) transition}

\begin{figure}
\includegraphics[width=\columnwidth]{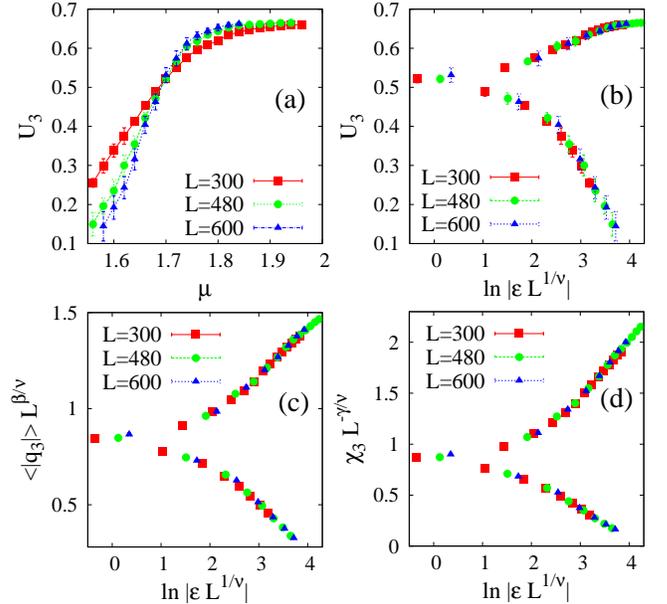}
\caption{The critical behavior near the N-C transition for rectangles of 
size $2 \times 15$ ($k=15/2$). (a) The data for Binder cumulant for 
different system sizes intersect at $\mu_{N-C}^c \approx 1.696$ 
($\rho^c_{N-C}\approx 0.75$). The data for different $L$ near the 
N-C transition for (b) Binder cumulant, (c) order parameter, and (d) 
second moment of the order parameter collapse onto a single curve when 
scaled as in Eq.~(\ref{eq:scaling}) with the Ising exponents 
$\beta/\nu=1/8$, $\gamma/\nu=7/4$, and $\nu = 1 $.}
\label{fig:scaling_NC_ch05}
\end{figure}

We study the N-C transition for rectangles of size $2 
\times 15$ using the order parameter $q_3$. When the system makes
a transition from the N phase with horizontal (vertical) orientation to the 
C phase, the symmetry between even and odd rows (columns) is broken. 
From symmetry
considerations, we expect the N-C transition to be in the Ising 
universality class. The data for $U_3$ for different system 
sizes intersect at $\mu=\mu_{N-C}^c\approx 1.696$ [see 
Fig.~\ref{fig:scaling_NC_ch05}(a)], corresponding to  $\rho^c_{N-C} \approx 0.759$.
The data for $U_3$ [see 
Fig.~\ref{fig:scaling_NC_ch05}(b)], $\langle |q_3| \rangle$ [see 
Fig.~\ref{fig:scaling_NC_ch05}(c)], and $\chi_3$ [see 
Fig.~\ref{fig:scaling_NC_ch05}(d)] for different system sizes collapse 
onto a single curve when scaled as in 
Eq.~(\ref{eq:scaling}) with Ising exponents 
$\beta/\nu=1/8$, $\gamma/\nu=7/4$, and $\nu = 1 $. The N-C transition in the
system of $2 \times 14$ rectangles has also been shown to be in 
the Ising universality 
class~\cite{joyjit_rectangle}, and we expect the same for 
$k > 15/2$.

\subsection{\label{sec:IC_ch05}Isotropic--Columnar (I-C) transition}

\begin{figure}
\includegraphics[width=\columnwidth]{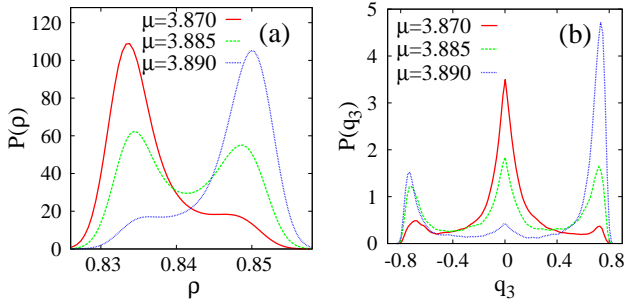}
\caption{Probability density function of (a) density 
$\rho$ and (b) order parameter $q_3$ for three values of $\mu$ near the 
I-C transition. The data are for rectangles of size $2\times 11$ and 
$L=440$.}
\label{fig:scaling_IC12_ch05}
\end{figure}

\begin{figure}
\includegraphics[width=\columnwidth]{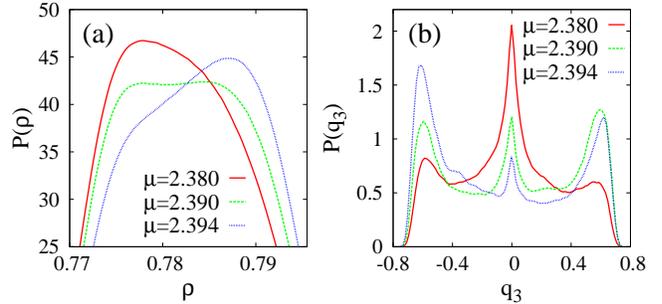}
\caption{Probability density function of (a) density 
$\rho$ and (b) order parameter $q_3$ for three values of $\mu$ near the 
I-C transition. The data are for rectangles of size $2\times 13$ and 
$L=416$.}
\label{fig:scaling_IC13_ch05} 
\end{figure}

The I-C transition occurs only for rectangles of size
$2\times 11$ ($k=11/2$) and $2 \times 13$ ($k=13/2$). The transition 
is best studied using the order parameter $q_3$. We find that the transition
is first order for both values of $k$. This may be established by numerically
calculating the probability density functions (pdf) $P(\rho)$ of the density $\rho$ and 
$P(q_3)$ of the order parameter $q_3$ for values of $\mu$ that are close 
to the transition point $\mu_{I-C}^{c} \approx 3.885$ for $k=11/2$ and 
$\mu \approx \mu_{I-C}^{c}=2.390$ for $k=13/2$. The pdf for $k=11/2$ and $13/2$ 
are shown in Fig.~\ref{fig:scaling_IC12_ch05} 
and Fig.~\ref{fig:scaling_IC13_ch05} respectively. In both the figures, the pdfs
for $q_3$ have three clear peaks at the transition point: the two peaks at 
$q_3\neq 0$ correspond to the symmetric
C phases and the one at $q_3=0$ to the I phase. 
The pdf for $\rho$ have two peaks of nearly equal height at the
transition point, though these peaks  are not clearly separated for $k=13/2$. The
difference in the peak positions is equal to the jump in the density across
the transition point and is shown by the shaded region in 
Fig.~\ref{fig:phase_diag_m2_ch05}.  We find that these peaks become sharper 
with increasing system size. These are clear signatures of a first order transition.

\section{Phase Diagram for $m \geq 3$}

We expect that the phase diagram for $m \geq 3$ and non-integer $k$ to be
qualitatively similar to that for $m=2$ with three entropy driven
transitions for large $k$. The HD phase is a disordered or 
sublattice phase depending on whether the length and width of the rectangles are
mutually prime or not. For $m=3$, we  determine the
the minimum value of $k$ beyond which the C and N phases exist. 
There are no transitions for $k\leq 11/3$. 
We find that the C phase exists only for  $k \geq 13/3$, while
the N phase exists only for $k \geq 22/3$. When $m=3$, the C phase 
has a 6-fold
symmetry, and thus with analogy with  Potts model, we expect a first order transition.
For rectangles with $k=13/3$, we  confirm that the I-C phase transition is first
order in nature. The pdfs for $\rho$ and $q_3$ behave similarly to that
for the case of $m=2$. All the other 
transitions are also expected to be first order except the I-N transition, as 
seen for integer $k$~\cite{joyjit_rectangle}.

\section{\label{sec:dis_ch05}Summary and Discussions}

In this paper, we obtained numerically the phase 
diagram of the system of hard rectangles of size $m\times mk$ with 
non-integer aspect ratio $k$. As for integer $k$, the system may exist in 
four different phases: isotropic, nematic, columnar or high density phase. 
For integer $k$, 
the high density phase is a solid-like sublattice phase. However, when
$k$ is a non-integer, the high density phase is a disordered phase when
the length and width of the rectangles are 
mutually prime. The phase diagram for large $m$ is expected to be qualitatively 
similar to that for $m=2$. The isotropic-nematic transition will be in
the Ising universality class for all $m$. However, all other transitions are expected 
to be first order.

The N phase  is found to exist only when  $k \geq 15/2$ for
$m=2$ and when $k\geq 22/3$ for $m=3$. For integer $k$, the N phase
exists only when $k\geq 7$ for $m=1,2,3$~\cite{ghosh2007,joyjit_rectangle}. These
different lower bounds may be combined to give tighter bounds for $k_{min}^{I-N}$, the
smallest  value of $k$ beyond which the N phase exists. 
We conclude that $20/3< k_{min}^{I-N} \leq 7$.

The bounds for $k_{min}^{I-C}$, the minimum value of $k$
beyond which the C phase exists, are not so clear. We find that
the C phase exists when $k \geq 11/2$ for $m=2$ and when $k \geq 13/3$ for
$m=3$. On the other hand, for integer $k$, the C phase exists for 
$k \geq 4$ for $m=2$ and when $k \geq 2 $ for $m=3$. Thus, unlike for the
N phase,  $k_{min}^{I-C}$ depends both on $m$ and whether $k$ is a integer
or not, and it is not possible to combine the bounds  only in terms of 
$m$.

The phase diagram when the aspect ratio of the rectangles
is irrational remains an open question.  In
this case, it has been conjectured that there could be more
transitions at densities close to full packing, when the disordered phase will
become unstable to a nematic or columnar phases~\cite{ghosh2007}. 
This question, as well as finding tighter bounds on $k$ for existence
of different phases are best answered by obtaining the phase diagram of 
hard rectangles with restricted orientation in two dimensional
continuum, when the aspect ratio may be continuously tuned.

\begin{acknowledgments}
The simulations were carried out on the supercomputing
machine Annapurna at The Institute of Mathematical Sciences. 
\end{acknowledgments}

%\bibliographystyle{eplbib}
%\bibliography{ref01}

\end{document}